\documentclass[12pt,letterpaper]{article}
\usepackage[a4paper, total={7in, 10in}]{geometry}

\usepackage{graphicx}
\usepackage{helvet}
\usepackage{authblk}
\usepackage{hyperref}
\usepackage{amsmath}
\usepackage{amssymb}
\usepackage{orcidlink}
\usepackage[super,comma,sort&compress]
   {natbib}

\graphicspath{{./Fig/}}
\makeatletter
\renewcommand{\maketitle}{\bgroup\setlength{\parindent}{0pt}
\begin{flushleft}
  \textbf{\@title}

  \@author
\end{flushleft}\egroup}
\makeatother

\title{Lithium-ion battery degradation: Introducing the concept of reservoirs to design for lifetime}
\date{}

\author[1,2\orcidlink{0000-0002-5082-7562}]{Mohammed Asheruddin Nazeeruddin}
\author[1\orcidlink{0000-0003-0965-8952}]{Ruihe Li}
\author[1\orcidlink{0000-0003-3141-1657}]{Simon E. J. O'Kane }
\author[1,2\orcidlink{0000-0003-1641-3371}]{Monica Marinescu}
\author[1,2\orcidlink{0000-0003-1324-8366}]{ Gregory J. Offer}

\affil[1]{Department of Mechanical Engineering, Imperial College London, London SW7 2AZ, United Kingdom}
\affil[2]{The Faraday Institution, Quad One, Becquerel Avenue, Harwell Campus, Didcot OX11 0RA, United Kingdom}

\affil[*]{Correspondence: mnazeeru@ic.ac.uk}

\begin{document}

\maketitle
\section*{Context \& scale}

Lithium-ion batteries (LIBs) are the backbone of electrified transport and grid-scale energy storage. Commercial designs often target energy densities of 250–300 Wh/kg, yet these gains are increasingly undermined by limitations in service life—with capacity fade often exceeding 20\% within 500–1,000 cycles under realistic charging, discharging, and thermal conditions. Despite the critical importance of long-term performance, battery cells are still largely designed around beginning-of-life optimization, emphasizing metrics such as initial capacity or energy density.

However, a growing body of evidence shows that even modest changes in design—such as electrode porosity or lithium excess—and in use conditions—such as charge rate or depth of discharge—can affect the dominant degradation pathway. These pathways can determine whether a cell fails via lithium inventory loss, active material isolation, or electrolyte dry-out and are rarely accounted for during the design stage. Most models employed in cell development treat degradation mechanisms in isolation, capturing individual effects, but failing to reveal the complex coupling that ultimately dictate aging behavior.

This work introduces a unifying design philosophy: reframing the key constituents of a cell — lithium, porosity, and electrolyte volume — as finite, coupled reservoirs that are progressively consumed through interacting degradation processes. Each reservoir has a size defined by cell design and a consumption rate governed by operating conditions. When any one reservoir is depleted, performance is compromised and capacity loss accelerates. By simulating reservoir depletion trajectories across a range of design and use scenarios, this framework enables a predictive, lifetime-aware approach to battery engineering. It allows designers to anticipate how a cell will age under realistic constraints—and to tailor cell architecture for specific applications, whether for fast-charging electric vehicles or long-duration storage systems.

\section*{SUMMARY}

Designing lithium-ion batteries for long service life remains a challenge, as most cells are optimized for beginning-of-life metrics such as energy density, often overlooking how design and operating conditions shape degradation. This work introduces a degradation-aware design framework built around finite, interacting reservoirs—lithium, porosity, and electrolyte—that are depleted over time by coupled degradation processes.

We extend a physics-based Doyle–Fuller–Newman model to include validated mechanisms — SEI growth, lithium plating, cracking, and solvent dry-out — and simulate how small design changes impact lifetime. Across 1,000+ cycles, we find that increasing electrolyte volume by just 1\% or porosity by 5\% can extend service life by over 30\% without significantly affecting cell energy density. However, lithium excess, while boosting initial capacity, can accelerate failure if not supported by sufficient structural or ionic buffers.

Importantly, we show that interaction between reservoirs is crucial to optimal design; multi-reservoir tuning yields either synergistic benefits or compound failures, depending on operating conditions. We also quantify how C-rate and operating temperature influence degradation pathways, emphasizing the need for co-optimized design and usage profiles.

By reframing degradation as a problem of managing finite internal reservoirs, this work offers a predictive and mechanistic foundation for designing LIBs that balance energy, durability, and application-specific needs.

\section*{KEYWORDS}

lithium-ion batteries, battery degradation, reservoir model, battery lifetime, loss of lithium inventory (LLI), loss of active material (LAM), electrolyte depletion, electrochemical modeling, battery design optimization, aging mechanisms

\section*{INTRODUCTION}

Lithium-ion batteries (LIBs) have emerged as the cornerstone of modern energy storage, powering everything from portable electronics to electric vehicles. Their unmatched combination of high energy density, stable cycling behavior, and declining cost has driven a global transition toward electrified transportation. As Goodenough observed, LIBs represent ``the most impressive success story of modern electrochemistry in the last two decades'' \cite{goodenough2014electrochemical}. Alongside rapid materials innovation, the maturation of physics-based modeling has profoundly reshaped battery research and development. Models grounded in the porous electrode theory introduced by Newman and co-workers\cite{doyle1993modeling, delacourt2012life} now enable simulation-based design and optimization, allowing researchers to probe internal dynamics, reduce experimental burden, and accelerate performance tuning \cite{birkl2017degradation}.

Early modeling efforts demonstrated how electrode parameters such as thickness, porosity, and active material loading influence initial performance metrics. For instance, it has been shown that increasing electrode thickness can enhance capacity but degrades rate performance---highlighting trade-offs that define conventional cell design \cite{yu2012investigation, heubner2019understanding}. However, despite their power, these models are most commonly deployed to optimize beginning-of-life (BoL) performance under idealized conditions\cite{etacheri2011challenges, cai2021boundary, appiah2016design, liu2022np, jagfeld2023aging, campbell2019optimising,wang2022liquid, liu2021electrolyte, weng2021predicting}, rather than to forecast and mitigate end-of-life (EoL) degradation\cite{sieg2020local,hosono2009synthesis,adam2020fast,liu2017optimal,grugeon2009combining,schmidt2019power,jha2022modeling,klick2023influence}. In both academic and industrial settings, optimization routines typically focus on energy density or peak power of pristine cells, seldom accounting for how those designs evolve with cycling. Moreover, parameters are often tuned one at a time, ignoring coupled effects and degradation pathways that emerge over hundreds or thousands of cycles \cite{rhyu2025systematic,pannala2024consistently}.

This BoL-centric paradigm overlooks the complex interplay of aging phenomena that govern long-term cyclability. Crucial mechanisms such as solid--electrolyte interphase (SEI) growth, lithium plating, electrolyte decomposition, and structural disconnection of active materials are largely absent from standard design pipelines\cite{weng2023modeling}. As a result, a cell that appears optimal in its fresh state may suffer accelerated degradation or catastrophic failure under realistic use conditions. While experimental studies have revealed these degradation processes in detail\cite{kirkaldy2024lithium, wildfeuer2023experimental}, their integration into design frameworks remains limited, underscoring the need for models that incorporate both performance and longevity\cite{stefanopoulou2023early}.

To address this gap, we present a degradation-aware modeling framework that integrates electrochemical performance with lifetime-limiting processes. Our approach leverages a coupled, multi-mechanism model capable of simulating SEI growth, lithium inventory loss, active material disconnection, porosity collapse, and electrolyte depletion within a unified physics-based formalism. Building on the Doyle--Fuller--Newman (DFN) model architecture, we incorporate validated extensions that capture key degradation pathways under realistic operating conditions. This enables direct simulation of capacity fade, mode-specific degradation, and cell failure as a function of design parameters and usage.

At the heart of this framework is a novel conceptual and quantitative tool: the reservoirs paradigm. While the idea of finite resources, such as cyclable lithium or active material, has appeared in experimental diagnostics \cite{movahedi2024extra, campbell2019optimising, beck2024improved, sohaib2025analysis, montaru2022calendar, dubarry2022perspective, pannala2022methodology}, it has not been systematically integrated into model-based design. We extend this concept into a multi-reservoir architecture that identifies and tracks five critical, depletable resources within the cell -- (1) Cyclable lithium, (2,3) Active material in the positive and negative electrode, (4) Electrolyte volume and (5) Electrode porosity. Each reservoir is treated as a finite resource, subject to progressive depletion, driven by mechanistically coupled degradation processes. For example, SEI formation consumes lithium and electrolyte while also contributing to pore clogging. Electrolyte oxidation diminishes ionic conductivity, raising overpotentials and further accelerating SEI growth and active material isolation. This feedback defines nonlinear depletion dynamics, in which multiple reservoirs interact, potentially triggering abrupt failure when any one reservoir reaches a critical threshold.

This framework provides a powerful abstraction for understanding why different cell designs fail in different ways. For a cell that is fast-charged, for instance, the cycle life may be limited by lithium plating and rapid lithium inventory loss; a cell that is operated at relatively high temperatures may degrade through solvent depletion and pore collapse. End-of-life, in this view, corresponds to terminal depletion of a limiting reservoir. This insight aligns with empirical findings, such as sudden capacity loss after dry-out\cite{li2023lithium1} , or early failure due to excessive lithium consumption \cite{von2019modeling}.

The reservoirs model enables rational design strategies aimed not merely at maximizing BoL performance but at balancing reservoir depletion rates to extend service life. For example, adding electrolyte increases the solvent reservoir, mitigating dry-out. Adjusting electrode porosity improves ionic accessibility and mechanical resilience. Tuning electrode formulation can slow SEI growth, preserving lithium inventory. Within our framework, these design choices are represented as reservoir sizing decisions that affect degradation trajectories in quantifiable ways.

Crucially, this approach also allows us to explore how coupled degradation pathways affect reservoir dynamics. Because each reservoir is linked through shared processes---such as SEI growth depleting lithium, electrolyte, and pore volume---the framework can identify which combinations of design parameters yield balanced, long-lived operation, and which create bottlenecks or compound failure loops.

In this study, we introduce and formalize the reservoirs concept, implement it within a validated physics-based modeling framework, and evaluate its implications through targeted simulations. We begin by isolating each reservoir---lithium inventory, porosity, and electrolyte volume---to understand its individual role in shaping degradation and end-of-life. We then investigate multi-reservoir tuning, analyzing how concurrent changes in multiple reservoirs interact and affect performance. Finally, we discuss how reservoir-sensitive design maps onto different real-world applications, such as fast-charging systems or long-duration storage.

This work offers a new paradigm for lithium-ion battery design: one that is degradation-aware, mechanistically grounded, and quantitatively predictive. By reframing cell components as finite, interacting reservoirs, we provide a unified methodology for extending battery lifetime through intentional design---empowering engineers to build cells that not only perform well initially but also age gracefully under realistic use.

\section*{Results and Discussion}

\subsection*{Effect of Individual Reservoirs}

Three reservoirs are considered  that can be controlled during the design process of a battery cell: lithium inventory, porosity, and electrolyte volume. In this section, the effect of independently tuning (increasing/decreasing) the size of each reservoir on battery performance is analyzed. The primary metric used to evaluate performance is number of cycles completed until 80\% State of Health (SoH)---i.e., until the battery capacity decreases to 80\% of its initial value. A cell with the longest cycle life to this SoH threshold is considered superior. However, it is important to emphasize that battery cycle life is just one example of a targeted performance metric. The modeling-aided battery design and operation approach proposed in this work can also be applied to optimize other performance criteria, such as energy throughput, reduced lithium plating, minimized SEI growth, and more.

Battery service life is estimated based on an assumed standard cycling protocol. A 5 Ah cell undergoes daily cycling at a rate of 10 Ah per day, equivalent to two full charge-discharge cycles. Over a year, this results in an accumulated charge throughput of approximately 3.65 kAh. The total charge throughput at which the battery reaches 80\% SoH is determined from model predictions. The service life, expressed in years, is then estimated by dividing this charge throughput by the annual cycling throughput. This methodology ensures a standardized and comparable measure of service life across different reservoir configurations.

In the following subsections, the impact of changing individual reservoir sizes---lithium, porosity, and electrolyte---on battery degradation is evaluated. A comparative analysis of battery lifetime as a function of reservoir size is presented, alongside traditional degradation modes such as Loss of Lithium Inventory (LLI) and Loss of Active Material (LAM), to explain the underlying mechanisms behind the observed trends. This analysis provides insights into how different degradation pathways dominate depending on reservoir availability and depletion rates, ultimately guiding design strategies for improving lithium-ion battery performance.

\subsubsection*{Lithium Reservoir}

The role of lithium inventory in shaping long-term degradation was evaluated by systematically tuning the initial cyclable lithium content in the negative electrode across four cell configurations: a baseline 5.00 Ah cell; a lithium-enriched design with +0.26 Ah (5.26 Ah); a lithium-deficient design with -0.26 Ah (4.74 Ah); and a lithium-enriched variant with modified positive electrode thickness to maintain electrode balancing. All simulations used the same electrochemical–mechanical–thermal model, incorporating interfacial side reactions, stress-driven active material loss, partially reversible lithium plating, and solvent dry-out.

\begin{figure}
  \centering
  \includegraphics[width=\linewidth]{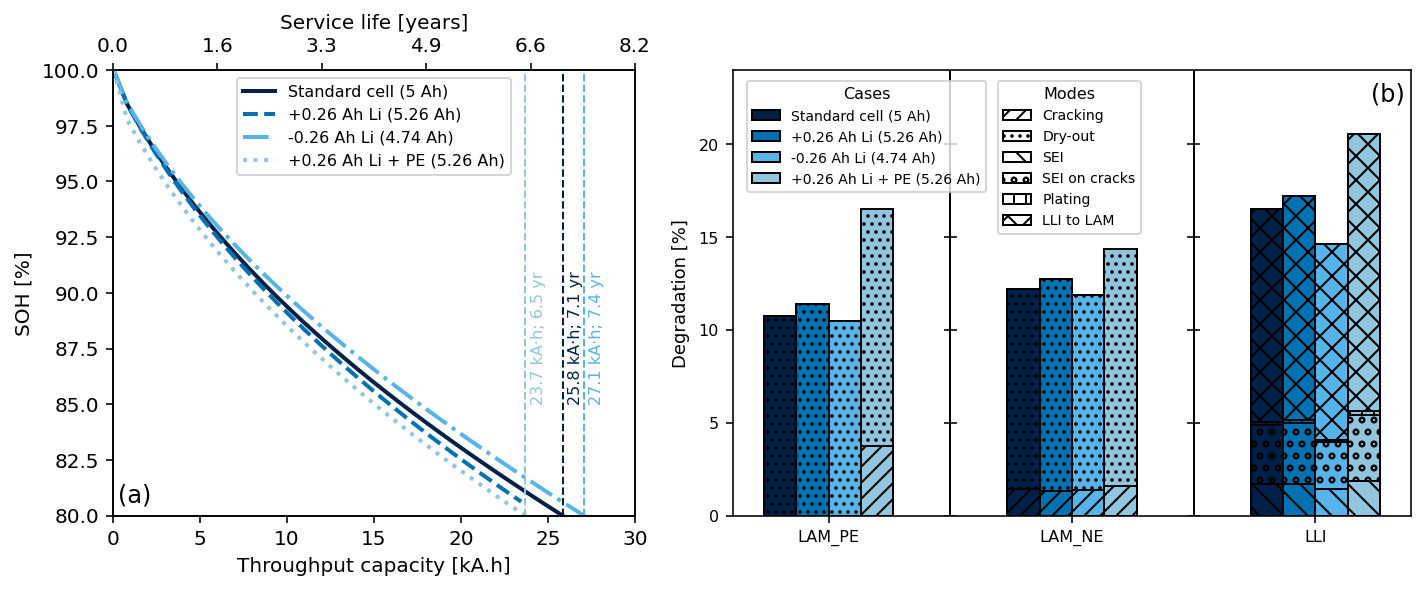}
  \caption{Effect of lithium reservoir (a) State-of-health (SoH) versus charge throughput for four lithium reservoir configurations: standard cell (5.00 Ah), +0.26 Ah Li (N/P = 0.82), ${-}0.26$ Ah Li (N/P = 0.75), and +0.26 Ah Li with adjusted positive electrode to retain N/P = 0.78. Service life is defined as the point at which SoH drops to 80\%. (b) Degradation mode breakdown at the 5-year mark (18.25 kAh), showing total loss of lithium inventory (LLI), loss of active material in the negative and positive electrodes (LAM\_NE, LAM\_PE), and their mechanistic contributors: LAM from cracking and dry-out, and LLI from SEI formation, lithium plating, and active material isolation.}
  \label{fig1}
\end{figure}

Figure \ref{fig1}a shows that the impact of lithium reservoir tuning is highly nonlinear. The lithium-deficient cell exhibited the longest service life, maintaining 84.64\% SoH at 5 years (18.25 kAh) and reaching 80\% SoH only after 27.07 kAh (7.42 years). The baseline 5.00 Ah cell reached 80\% SoH after 25.85 kAh (7.08 years), while the +0.26 Ah lithium and +0.26 Ah + PE configurations degraded faster, reaching 80\% SoH after 24.19 kAh (6.62 years) and 23.71 kAh (6.49 years), respectively. These results challenge the intuitive assumption that adding lithium uniformly enhances cell life.

Figure \ref{fig1}b provides a breakdown of degradation modes at the 5-year mark (18.25 kAh), near end-of-life for all configurations. Three macro-level degradation metrics—loss of lithium inventory (LLI), loss of active material in the negative electrode (LAM\_NE), and in the positive electrode (LAM\_PE)—are decomposed into mechanistic subcomponents: SEI growth across the bulk and on cracks, lithium plating, lithium isolation (LLI due to LAM), and LAM due to dry-out and cracking.

A key caveat must be addressed before interpreting these values. In the solvent model by Li et al. \cite{li2022modelling}, dry-out is triggered by EC solvent consumption during SEI growth in the negative electrode, leading to porosity loss. Once porosity falls below a critical threshold, the simulation removes the entire through-plane path—including the PE, separator, and NE—for numerical stability. While physically, dry-out is localized to the NE, this implementation results in artificial inactivation of corresponding PE regions. Consequently, post-processed LAM\_PE includes capacity loss not due to actual structural degradation in the PE but due to NE-originated dry-out. This modeling artifact is critical to interpret in contexts where apparent PE degradation appears despite zero physically modeled PE cracking.

In the baseline 5.00 Ah cell, total LLI is 16.52\%, dominated by lithium isolation (11.46\%), with moderate SEI formation (1.68\% bulk, 3.21\% on cracks) and negligible plating (0.17\%). LAM\_NE totals 12.22\%, with 10.77\% from dry-out and 1.45\% from cracking. The PE shows 10.77\% LAM—equal to the dry-out in the NE—despite no modeled PE cracking. This is a direct consequence of the symmetric removal of through-plane paths and highlights the importance of distinguishing real degradation from numerical inactivation. Physically, the PE remains intact.

Adding 0.26 Ah lithium without rebalancing the PE accelerates degradation. LLI increases to 17.22\%, with lithium isolation rising to 12.07\%, and slightly higher SEI growth (1.71\% bulk, 3.27\% on cracks). LAM\_NE rises to 12.73\%, again dominated by dry-out. The apparent LAM\_PE is 11.40\%, again attributable to NE dry-out. The added lithium allows the NE to reach deeper lithiation (closer to 0 V vs. Li/Li$^+$), increasing interfacial stress and promoting side reactions. Although plating remains minor (0.18\%), the elevated utilization of lithium intensifies irreversible lithium disconnection and porosity loss, leading to earlier reservoir exhaustion.

Conversely, the lithium-deficient cell exhibits the most favorable degradation profile. LLI drops to 14.20\% (with only 10.57\% from lithium isolation), SEI formation is reduced (1.42\% bulk, 2.53\% on cracks), and plating is minimal (0.12\%). LAM\_NE also decreases to 9.58\%, primarily through suppressed dry-out (10.49\%) and only a modest rise in cracking (1.97\%). This case operates in a narrower potential window, avoiding the deep lithiation of the NE that triggers structural and interfacial degradation. Despite the smaller reservoir, this cell degrades more slowly, and the reduced initial capacity is offset by longer service life.

The most aggressive degradation occurs in the +0.26 Ah + PE configuration. LLI peaks at 20.53\%, with 14.89\% from lithium isolation and the highest SEI growth (1.87\% bulk, 3.57\% on cracks). LAM\_NE rises to 14.38\%, and for the first time, true PE damage appears: 3.75\% LAM\_to\_Crack\_PE. The total LAM\_PE is 16.52\%, with only part attributable to actual cracking. Here, both electrodes are driven into extreme potential regimes—deep lithiation of the NE and high-voltage delithiation of the PE—leading to simultaneous structural degradation, particularly at the cathode. The elevated lithium inventory and matched PE thickness expand the electrode potential windows beyond mechanically stable limits, triggering stress-driven cracking in both electrodes.

These findings collectively highlight that lithium reservoir tuning is far from a one-dimensional design lever. While a larger reservoir can offset early-cycle inefficiencies and marginally improve initial capacity, it often promotes deeper lithiation of the negative electrode, increasing mechanical strain and accelerating both lithium isolation and SEI turnover—ultimately leading to faster capacity fade. In contrast, lithium-limited designs, though initially constrained in energy density, restrict the accessible lithiation window and thereby avoid high-stress regimes associated with cracking and lithium plating. This trade-off results in more stable long-term performance and delayed reservoir depletion. Notably, simply rebalancing electrode capacities to accommodate added lithium, such as by thickening the positive electrode, does not prevent degradation; in fact, it can push both electrodes into extreme potential windows, activating new degradation pathways like positive electrode cracking that were absent in the baseline configuration. Therefore, reservoir optimization must be pursued as part of a broader, multi-dimensional strategy that jointly considers electrode mechanics, electrolyte stability, interfacial reactivity, and system-level cycling conditions. Only when these interdependencies are understood and integrated can the lithium reservoir be effectively leveraged to improve durability, not just initial performance.

In summary, lithium reservoir tuning is not a matter of “more is better.” Its impact is context-specific and dependent on how the extra lithium interacts with the electrode mechanics, potential windows, and electrolyte stability. Reservoir-aware design must consider both the quantity of lithium and the conditions under which it will be cycled. Models that account for physical degradation mechanisms, but are mindful of implementation artifacts, can offer powerful insight—but their outputs must be interpreted with care, especially when attributing causality to structural degradation. Only then can we develop lithium-ion cells optimized not just for initial capacity, but for endurance under real-world conditions.

\subsubsection*{Porosity Reservoir}

Electrode porosity is a foundational design parameter in lithium-ion cells—governing ionic transport, electrolyte accessibility, and structural resilience. Within the reservoir-based framework developed in this work, porosity is treated as a finite resource that depletes over cycling, triggering cascading degradation. To isolate the role of this structural reservoir, we consider a set of cells with identical active material loading but varying electrode thickness ratios ($R_{\mathrm{thickness}}$), which directly alter porosity through calendering. Importantly, because calendering compresses the electrode structure without changing total active material, energy density changes arise purely from volumetric compaction, not from differences in electrochemical capacity. Increasing $R_{\mathrm{thickness}}$ (i.e., reducing calendering pressure) results in higher porosity and lower volumetric energy density, while decreasing $R_{\mathrm{thickness}}$ increases compaction and reduces pore volume. Specifically, the baseline cell has $R_{\mathrm{thickness}} = 1.00$ and porosity $\varepsilon = 0.222$; the compressed cell ($R_{\mathrm{thickness}} = 0.95$) has reduced porosity ($\varepsilon = 0.181$), and the expanded cell ($R_{\mathrm{thickness}} = 1.05$) maintains slightly higher porosity ($\varepsilon = 0.220$). These geometric changes—while modest—have profound implications for porosity reservoir depletion and its coupling with electrochemical and mechanical degradation.

\begin{figure}
  \centering
  \includegraphics[width=\linewidth]{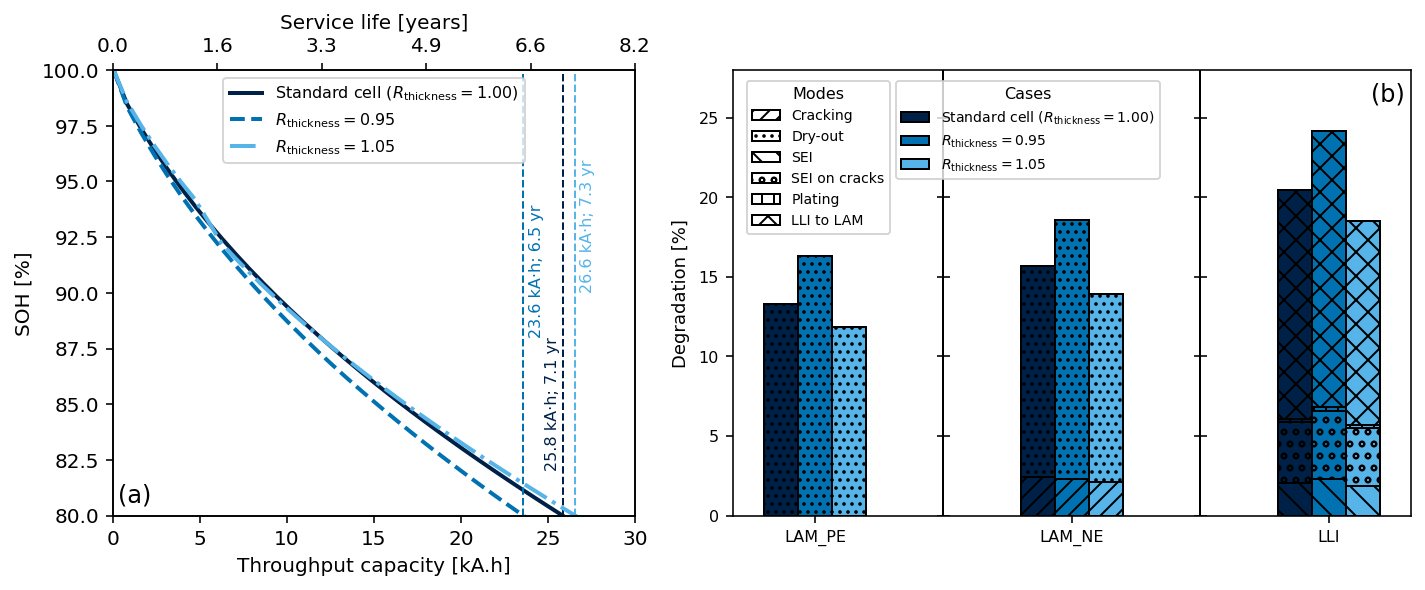}
  \caption{Effect of porosity reservoir tuning on battery performance. (a) State-of-health (SoH) versus charge throughput for varying electrode thickness ratios ($R_{thickness}$= 0.95, 1.00, 1.05). (b) Degradation mode breakdown at the 5-year mark (18.25 kAh), showing LLI and LAM contributions from SEI, plating, cracking, dry-out, and active material isolation.}
  \label{fig2}
\end{figure}

As shown in Figure \ref{fig2}a, compressing the electrode accelerates degradation and reduces service life. The compressed cell reaches 80\% state-of-health (SoH) after 23.59 kAh (6.46 years), compared to 25.85 kAh (7.08 years) for the baseline and 26.57 kAh (7.28 years) for the expanded configuration. While compaction improves volumetric energy density—a desirable attribute in many applications—it restricts electrolyte-filled pore volume, raises local current densities, and exacerbates mass transport limitations. These conditions drive the electrode toward premature reservoir depletion.

Figure \ref{fig2}b reveals the degradation landscape at a common throughput of 18.25 kAh, showing that loss of lithium inventory (LLI) rises significantly with compression—from 18.49\% in the expanded case to 24.16\% in the most compressed cell. The majority of this loss stems from lithium isolation due to active material disconnection (LLI due to LAM), which increases from 12.82\% to 17.35\% across the same range. This trend points to dry-out—not direct interfacial reactions—as the dominant degradation mechanism at low porosity.

In the model, dry-out arises from localized electrolyte depletion, driven primarily by SEI formation in the negative electrode. As the SEI consumes electrolyte solvent, local porosity declines. Once porosity falls below a critical threshold, ionic transport becomes unsustainable, and the region is deactivated—cut off from further electrochemical participation. This leads simultaneously to LLI, from trapped lithium, and to LAM, from loss of conductive pathways. While lithium plating also consumes electrolyte and lithium, its contribution remains minor (~0.20\%) across all configurations. Thus, the strong porosity dependence of dry-out-induced degradation reflects a coupled chain of events initiated by interfacial reactions and structurally amplified by compression.

Negative electrode degradation (LAM\_NE) exhibits a similar progression: increasing from 13.94\% in the expanded case to 18.60\% in the compressed cell. This increase is almost entirely due to dry-out, which rose from 11.84\% to 16.31\%, while cracking-related LAM remained relatively stable (2.10–2.42\%). This structural degradation does not stem from silicon expansion (as silicon is not included in this model), but from electrolyte starvation and pore collapse. Although the model reports nonzero LAM in the positive electrode (LAM\_PE), this is a numerical artifact of the dry-out implementation: once a region in the negative electrode reaches the porosity threshold, the full through-plane domain—including the PE and separator—is removed. Since the PE is not subject to mechanical or electrochemical degradation in this model, the reported LAM\_PE does not reflect true physical failure.

The expanded configuration ($R_{\mathrm{thickness}} = 1.05$) outperforms both other cases in durability, with the lowest LLI and LAM\_NE and the longest service life. However, this gain comes at the cost of reduced volumetric energy density. The degradation trends are also nonlinear: a ~9\% reduction in porosity (from 0.222 to 0.202) leads to a \~30\% increase in dry-out-induced LLI. This amplification illustrates the sensitivity of degradation to porosity tuning and the dangers of operating near structural and transport thresholds. While lower porosity may improve rate capability by reducing tortuosity, this short-term gain is often offset by accelerated depletion of the electrolyte and porosity reservoirs.

These findings underscore that porosity is not merely a passive structural parameter—it is an active, consumable reservoir whose exhaustion triggers electrochemical isolation and structural collapse. While high compression may be preferred for mobile or high-energy-density formats, it incurs a long-term penalty in terms of LLI, LAM, and service life. Conversely, expanded porosity configurations delay reservoir depletion and enhance cycling stability, making them better suited for applications such as stationary storage, where longevity outweighs footprint.

Crucially, these insights emerge only within a degradation-aware, reservoir-based modeling framework that tracks porosity depletion and its cascading effects. This approach reveals that porosity tuning cannot be evaluated in isolation. It must be co-optimized with lithium inventory, electrolyte capacity, and mechanical resilience. A holistic design strategy must account for this interdependence. It should target not only energy density or impedance, but also structural sustainability over the cell’s full lifecycle.

\subsubsection*{Electrolyte Reservoir}

Electrolyte volume represents a critical finite reservoir in lithium-ion cells. Over time, parasitic reactions— particularly solid-electrolyte interphase (SEI) formation and lithium plating—irreversibly consume electrolyte solvent. This not only reduces the bulk ionic conductivity but also leads to local electrolyte starvation, pore clogging, and accelerated degradation at electrode interfaces. In this study, we assess the system-wide consequences of minor variations in electrolyte excess ratio ($R_\mathrm{excess}$) on degradation dynamics and service life. Three cases were simulated: a baseline cell ($R_\mathrm{excess}$ = 1.00), an electrolyte-deficient configuration ($R_\mathrm{excess}$ = 0.99), and an electrolyte-excess design ($R_\mathrm{excess}$ = 1.01), with all other parameters held constant.

\begin{figure}
  \centering
  \includegraphics[width=\linewidth]{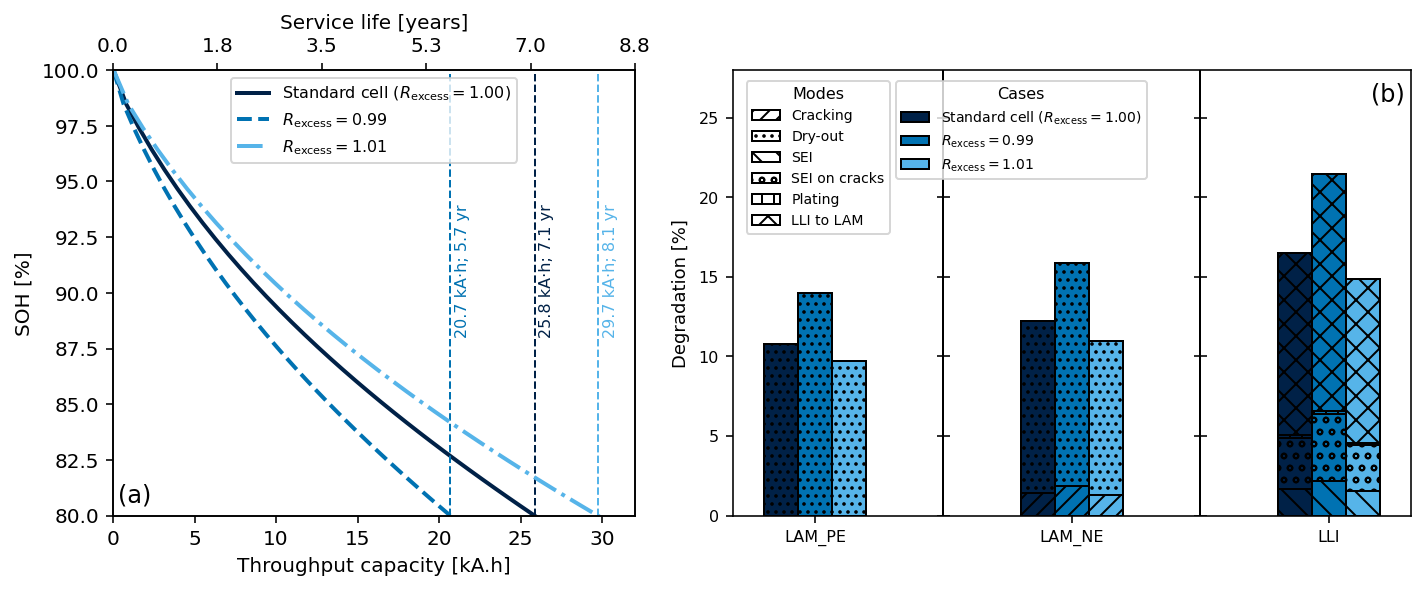}
  \caption{\textit{Effect of electrolyte reservoir size on cell performance and degradation.} (a) State-of-health (SoH) versus charge throughput for cells with varying electrolyte excess ratios ($R_{excess}$ = 0.99, 1.00, 1.01). (b) Degradation breakdown at the 5-year mark (18.25 kAh), showing contributions to loss of lithium inventory (LLI) and loss of active material in the negative and positive electrodes (LAM\_NE, LAM\_PE), including mechanisms such as SEI formation, plating, and dry-out.}
  \label{fig3}
\end{figure}

As shown in Figure~\ref{fig3}a, electrolyte availability exerts a pronounced and nonlinear effect on cell longevity. The electrolyte-deficient case reaches 80\% state-of-health (SoH) after only 20.68 kAh of cumulative throughput (5.67 years), compared to 25.85 kAh (7.08 years) for the baseline and 29.72 kAh (8.14 years) for the electrolyte-excess cell. Despite only a 2\% variation in initial electrolyte volume between the leanest and richest cases, the improvement in service life exceeds 44\%, underscoring the compounding effect of electrolyte availability on multiple coupled degradation pathways.

Figure~\ref{fig3}b shows that the total loss of lithium inventory (LLI) increases significantly in the electrolyte-deficient case, reaching 21.48\% by the end of 5 years, compared to 16.52\% and 14.87\% in the baseline and excess-electrolyte designs, respectively. Most of this increase stems from LLI due to active material isolation—14.90\% in the $R_\mathrm{excess}$ = 0.99 case—reflecting how electrolyte starvation accelerates structural and interfacial degradation. In contrast, direct contributions from SEI growth (2.19\%) and lithium plating (0.22\%) remain modest across all designs. These results suggest that solvent loss not only impairs ionic transport but also destabilizes electrode microstructures, compounding the effects of mechanical degradation.

Loss of active material in the negative electrode (LAM\textsubscript{NE}) follows a similar trend. The electrolyte - deficient cell exhibits 15.88\% LAM\textsubscript{NE}, compared to 12.22\% and 11.00\% for the baseline and electrolyte - excess configurations. The increase is largely driven by LAM due to dry-out, which rises from 10.77\% in the baseline to 14.00\% in the deficient cell. This reflects enhanced electrolyte depletion near the SEI layer, pore closure, and local overpotentials that trigger further cracking and interfacial failure. Even a modest electrolyte surplus ($R_\mathrm{excess}$ = 1.01) provides a buffer against these effects, lowering both dry-out and LAM\textsubscript{NE} while enabling more uniform ionic conduction.

The positive electrode also appears to exhibit a decline in capacity under electrolyte-deficient conditions, with LAM\textsubscript{PE} rising from 9.69\% ($R_\mathrm{excess}$ = 1.01) to 14.00\% ($R_\mathrm{excess}$ = 0.99). However, this must be interpreted with caution. In the modeling approach used here, once local porosity in the negative electrode drops below a critical threshold due to solvent loss, the model removes the entire through-plane path—including PE, separator, and NE—from the active domain. This can result in artificial increases in PE LAM even when no direct degradation mechanisms are active in the positive electrode. Since our model does not explicitly include electrolyte oxidation, transition metal dissolution, or surface reconstruction in the PE, the rise in LAM\textsubscript{PE} in lean-electrolyte conditions is best viewed as a model artifact rather than a physically resolved phenomenon.

Altogether, these results highlight how electrolyte depletion affects degradation in a coupled and system-wide manner. In the electrolyte-deficient cell, LLI, LAM\textsubscript{NE}, and LAM\textsubscript{PE} all rise concurrently, reflecting a regime of compounding degradation where solvent loss, interface breakdown, and structural failure reinforce one another. By contrast, the electrolyte-excess design suppresses all major degradation pathways, extending service life and stabilizing the cell against localized failure. Notably, even the small increase in initial solvent volume afforded by $R_\mathrm{excess}$ = 1.01 delays end-of-life by more than a full year.

From a design standpoint, these findings reinforce the importance of electrolyte reservoir tuning as a lever for balancing gravimetric/volumetric energy density against long-term durability. The electrolyte-excess configuration is particularly advantageous for applications requiring extended calendar life, such as stationary storage or electric vehicles under fast-charging protocols. Meanwhile, electrolyte-deficient designs may be justified in high-density formats but demand careful mitigation strategies, including reduced cycling rates, surface coatings, or electrolyte additives. Crucially, the reservoir-based modeling framework used in this work enables explicit quantification of electrolyte depletion and its impact on coupled degradation pathways, offering a foundation for targeted electrolyte optimization in degradation-aware battery design.

\subsection*{Effect of Multi-Reservoir Variations}

To assess the co-optimization potential of reservoir parameters, we systematically explored the effects of multi-reservoir tuning on service life and energy density, using the standard cell as the reference baseline. Each configuration varied lithium inventory (L), electrode porosity (P), and electrolyte volume (E) in controlled increments: $\pm0.26$ Ah for lithium, $\pm5\%$ for porosity (via calendering), and $\pm1\%$ for electrolyte. These perturbations were selected to reflect realistic engineering tolerances in advanced cell manufacturing. For each configuration, the relative improvement in performance was quantified using the improvement factor - the ratio of a given metric (service life, gravimetric energy density, or volumetric energy density) to that of the standard cell.

\begin{figure}
  \centering
  \includegraphics[width=0.8\linewidth]{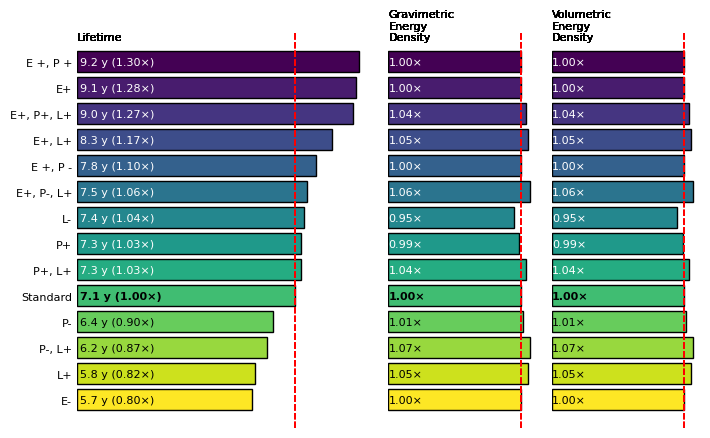}
  \caption{Relative improvement in service life and energy density for various multi-reservoir configurations, expressed as a ratio to the standard cell. Each bar represents a unique combination of lithium inventory (L), porosity (P), and electrolyte volume (E), with ‘+’ and ‘-’ indicating increases or decreases from baseline values. Left panel shows service life improvement; middle and right panels show corresponding changes in gravimetric and volumetric energy density.}
  \label{fig4}
\end{figure}

The results reveal a striking asymmetry: substantial increases in service life can be achieved with minimal or even positive changes in energy density. For example, the combined configuration L+P+E+ extended service life by $1.27\times$ while simultaneously improving both gravimetric and volumetric energy density by approximately 4\%. Even single-parameter enhancements such as E+ alone (increased electrolyte) or P+E+ (porosity and electrolyte excess) delivered $>$25\% service life extensions with negligible mass or volume penalties.

This phenomenon arises from the fact that energy density is relatively insensitive to small reservoir changes. For instance, a $\pm5\%$ change in porosity alters the electrode thickness but contributes only $\sim$1.3\% to the overall cell mass and volume, given that active material constitutes just $\sim$25\% of the full cell. Similarly, electrolyte volume changes - though crucial for long-term degradation - represent less than 0.1\% of the total mass due to the low density and fractional volume of the liquid phase. Even lithium inventory changes, while directly affecting capacity, remain diluted in the full-cell structure, causing small shifts in gravimetric or volumetric energy density. Consequently, modest increases in reservoir size do not impose significant penalties on energy storage metrics.

In contrast, service life is highly sensitive to reservoir depletion, due to nonlinear amplification of degradation once critical thresholds are crossed. Porosity and electrolyte volume directly influence ionic conductivity and wetting, and their early depletion accelerates transport limitations, electrode dry-out, and interfacial instability. Excess electrolyte helps buffer the cell against local starvation and supports uniform SEI growth, while increased porosity mitigates local overpotentials and mechanical damage by allowing more breathing room for electrode expansion. When combined with lithium over-stoichiometry (L+), which delays the onset of lithium inventory loss (LLI), these configurations act synergistically to suppress degradation across multiple pathways.

However, synergy is not guaranteed. Some configurations exhibit diminishing returns or even adverse effects when reservoirs are misaligned. Notably, L+ in isolation reduced service life to $0.82\times$ the baseline despite modest energy gains. This suggests that added lithium, without sufficient electrolyte or porosity, can accelerate degradation through excessive lithiation, increased strain, and localized plating. Similarly, porosity reduction (P$-$) yielded a 10\% drop in service life with minimal gain in volumetric energy density, reinforcing that aggressive compaction may compromise long-term stability more than it improves storage density.

Taken together, these findings highlight that reservoir tuning must be approached as a coupled, mechanism-aware optimization problem. The ability to extend service life by $>$30\% - and in some cases enhance energy density - demonstrates a powerful design lever. But the benefits are realized only when lithium, porosity, and electrolyte are co-optimized to avoid early depletion or isolated bottlenecks. Poorly coordinated reservoir changes can degrade both life and capacity, underscoring the importance of integrated design strategies.

This analysis affirms a central thesis of this work: battery degradation is a multidimensional, reservoir-mediated process. Engineering gains lie not in single-point optimizations but in tuning the collective resilience of the system. The reservoir-based framework introduced here offers a predictive and quantitative platform for such optimization, enabling rational design of high-performance lithium-ion batteries tailored to the evolving demands of electric mobility, fast charging, and long-duration grid storage.

\subsection*{Effect of Operating Conditions on Reservoir Dynamics}

This section quantitatively evaluates the influence of external operating conditions—specifically charge/discharge rate (C-rate/D-rate) and temperature—on reservoir exhaustion and degradation progression in lithium-ion cells. By fixing the reservoir configuration (baseline lithium inventory, porosity, and electrolyte volume), we isolate the impact of operational stressors on the rates and mechanisms of degradation.

\begin{figure}
  \centering
  \includegraphics[width=\linewidth]{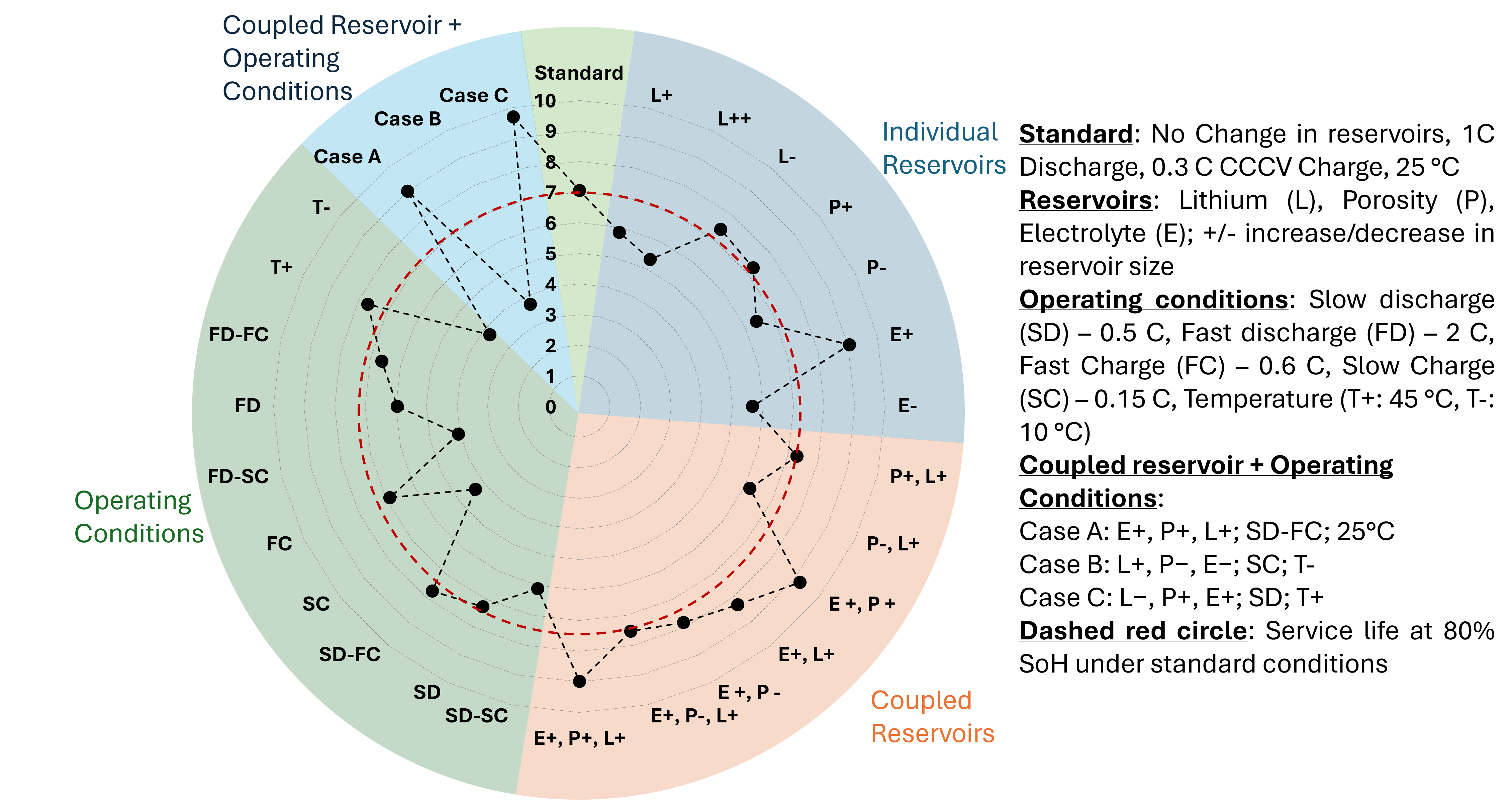}
  \caption{Service life at 80\% state of health (SoH) under varying reservoir sizes, coupled reservoir configurations, operating conditions, and their combined effects.}
  \label{fig5}
\end{figure}

\subsubsection*{Influence of C-Rate and D-rate on Reservoir Utilization and Degradation}

Figure \ref{fig5} illustrates the nuanced effects of varied charge (C-rate) and discharge (D-rate) rates on reservoir utilization and degradation dynamics. While the underlying reservoir structure remains unchanged, current profiles modulate degradation trajectories through their control over lithium distribution, interfacial conditions, and mechanical strain.

At a low-rate regime of 0.5D/0.15C, the cell surprisingly reaches end-of-life after just 6.11 years—shorter than several higher-rate scenarios. This counterintuitive result stems from slow charge kinetics, which induce steep lithium concentration gradients and localized overpotentials. These gradients trigger non-uniform SEI growth and mechanical stress hotspots, resulting in accelerated lithium inventory loss (LLI) and loss of active material (LAM) even in the absence of aggressive electrochemical loading.

Increasing the charge rate to 0.3C and 0.6C (with 0.5D fixed) enhances service life to 7.27 and 7.71 years, respectively. The improvement is attributed to faster lithium reinsertion and homogenized electrode utilization. As shown in Figure \ref{fig5}, these faster kinetics suppress parasitic side reactions and reduce interfacial resistance buildup, underscoring the value of balanced kinetic profiles.

At a higher discharge rate of 1D, service life deteriorates sharply to 4.35 years when paired with slow 0.15C charging. The mismatch in charge-discharge rates creates lithium saturation asymmetries, increasing the risk of plating and localized impedance rise. However, raising the charge rate to 0.6C mitigates these effects and extends life to 6.88 years by rebalancing electrode lithiation and improving charge recovery.

When subjected to even more aggressive 2D discharge, degradation intensifies. Life ranges from 4.06 years (0.15C) to 6.62 years (0.6C). Here, the dominant drivers are intercalation-induced stress, severe intra-particle concentration gradients, and mechanical fracturing — particularly in the negative electrode. While higher charge rates assist with ionic rebalancing, they cannot fully counteract the structural damage initiated during high-rate discharge.

These findings converge to identify a favorable operating window: discharge rates between 0.5D and 1D, paired with charge rates between 0.3C and 0.6C. In contrast, excessively slow charging (e.g., 0.15C) paradoxically accelerates degradation. Though seemingly gentle, such profiles prolong high SoC residence time and operation near anode equilibrium potential — conditions that thermodynamically favor SEI growth and lithium plating. Even modest kinetic asymmetries can thus trigger nonlinear amplification of irreversible degradation.

In summary, charge and discharge current profiles serve as levers for modulating the rate and spatial distribution of reservoir depletion. Optimal C/D-rate coordination—even under fixed reservoir configurations—can substantially delay capacity fade, highlighting the critical role of electrochemical protocol design in degradation-aware battery engineering.

\subsubsection*{Influence of Temperature on Reservoir Depletion Rates}

Temperature is a master variable in lithium-ion cell performance, modulating ionic conductivity, interfacial kinetics, and solid-state diffusion. Simulations show that even modest changes in temperature drastically alter the exhaustion rates of lithium, electrolyte, and porosity reservoirs.

At $10^\circ$C, the cell reaches end-of-life in just 3.75 years. The sharp decline in performance is driven by inhibited transport — both ionic and solid-state — which results in steep potential gradients and sluggish lithiation kinetics. These conditions promote lithium plating, uneven SEI growth, and intercalation asymmetries, leading to early LLI and stress-driven LAM. Additionally, low exchange currents increase charge-transfer resistance, exacerbating interfacial losses, especially during post-discharge recovery.

At $40^\circ$C, by contrast, service life extends to 7.68 years, surpassing the 7.1-year baseline at $25^\circ$C. Elevated temperature enhances lithium diffusivity and exchange current densities, facilitating more uniform reservoir usage and suppressing spatial degradation. However, the benefits of higher temperature come at a cost: the same thermal activation that improves kinetics also accelerates side reactions, such as SEI thickening, electrolyte oxidation, and gas evolution.

Thus, temperature exhibits a dual role—enabling performance and degrading stability. While moderate thermal elevation can enhance lifespan, especially under high-rate applications, sustained exposure above $40^\circ$C without active cooling leads to accelerated reservoir depletion. Effective battery design must therefore incorporate thermal management systems that maintain operation within kinetically favorable but chemically safe windows.

\subsubsection*{Interactions Between Operating Conditions and Reservoir Dynamics}

To reveal synergistic or antagonistic interactions between structure and operation, we analyze three exemplar cases:

\textbf{Case A: E+, P+, L+ at 0.5C/0.6C, $25^\circ$C} — This configuration combines structural resilience with thermal and kinetic stability. Enhanced porosity and electrolyte content (P+, E+) minimize ionic resistance and thermal hotspots, while lithium overcapacity (L+) delays LLI onset. Moderate-fast charging avoids low-anode-potential regimes, and elevated porosity softens mechanical stress accumulation. Together, these factors suppress plating and cracking, yielding a projected service life of 9.0 years.

\textbf{Case B: L+, P$^{-}$, E$^{-}$ at 1C/0.15C, $10^\circ$C} — A textbook case of degradation compounding. High lithium inventory without structural support (P$^{-}$, E$^{-}$) and under cold, slow-charging conditions leads to lithium saturation, severe overpotentials, and interfacial breakdown. Low temperature exacerbates charge-transfer bottlenecks, and deficient porosity amplifies cracking. The result: only 3.6 years of service life, dominated by early LLI and rapid impedance rise.

\textbf{Case C: L$^{-}$, P+, E+ at 0.5C/0.3C, $45^\circ$C} — An example of lifetime optimization through chemical moderation and structural enhancement. Reduced lithium (L$^{-}$) limits SEI formation and excess capacity loss, while structural surplus (P+, E+) promotes uniform transport and thermal buffering. Elevated temperature accelerates kinetics without breaching degradation thresholds. This configuration achieves 9.7 years of life, albeit with a modest trade-off in total energy capacity.

These scenarios underscore that reservoir longevity is not a static material property but an emergent result of multidimensional interactions. Even with fixed material sets, operational tuning can prolong or curtail lifespan by over 40\%.

\textbf{Application-Specific Implications:}
\begin{itemize}
\item \textbf{Electric Vehicles (EVs):} Require aggressive charge rates (5--10C), high throughput, and wide thermal tolerance. Structural enhancements (E+, P+) buffer against plating and thermal gradients. Preheating cells to $>\$25\$^\circ$C prior to fast charging minimizes parasitic losses. While lithium surplus (L+) helps retain capacity, it must be carefully balanced with robust BMS controls.
\item \textbf{Stationary Energy Storage Systems (ESS):} Prioritize calendar life and reliability over energy density. Operation at mild C-rates ($\le$0.5C) and ambient temperatures is optimal. Lithium moderation (L$^{-}$) and structural buffering (E+, P+) enhance passive stability, reduce side reactions, and maintain long-term impedance control.
\item \textbf{Consumer Electronics:} Constrained by size and weight, often featuring P$^{-}$ and E$^{-}$ designs. These cells are vulnerable to local dry-out and plating under aggressive use. Mitigating degradation requires software-based charge throttling (e.g., SoC caps), moderate C-rates (0.3C-0.5C), and reduced lithium loading (L$^{-}$) to avoid SEI overgrowth.
\end{itemize}

These findings make a compelling case for application-informed, degradation-aware battery design. By simulating the interplay between reservoir capacities and operational stressors, we reveal how electrochemical protocols—not just material choices—govern long-term cell stability. The emergent nonlinearities in degradation dynamics demand a shift from rule-of-thumb engineering to predictive, model-based design. Reservoir exhaustion is not simply a function of time or cycles, but a dynamic consequence of coupled thermal, mechanical, and chemical processes. Unlocking the full potential of lithium-ion systems will require holistic co-optimization of structure, chemistry, and operation — well within reach through advanced modeling frameworks.

\section*{Conclusions}

This study introduces a degradation-aware design framework for lithium-ion batteries built around the concept of reservoirs—finite, interacting stores of cyclable lithium, active material, electrolyte, and porosity that collectively dictate the trajectory of cell aging. By embedding this paradigm within a physics-based electrochemical model that incorporates validated degradation mechanisms—including SEI growth, lithium plating, particle cracking, and electrolyte dry-out—we reveal how reservoir depletion drives nonlinear, coupled degradation pathways that often yield counterintuitive outcomes.

Through systematic tuning of individual and combined reservoirs, we uncover key design insights. Modest increases in electrolyte volume or porosity can extend service life by over 30\%, with minimal penalty to gravimetric or volumetric energy density. In contrast, increasing lithium inventory without sufficient structural or ionic buffering can accelerate degradation via early onset of plating or lithium isolation. These results highlight that no reservoir acts in isolation—and that well-aligned configurations can synergistically delay failure, while poorly matched choices can amplify degradation across multiple fronts.

Furthermore, we show that operating conditions such as charge/discharge rates and temperature play a decisive role in modulating reservoir exhaustion. Even with fixed material configurations, variations in thermal and electrochemical stress reshape degradation kinetics. For instance, slow charging may unintentionally increase time spent in parasitic regimes, while mild thermal elevation can stabilize interfacial reactions when properly managed. These findings emphasize that lifetime optimization is not merely a material challenge—but an operational one as well.

Taken together, this work reframes lithium-ion battery design as a problem of intelligently allocating and preserving finite degradation-sensitive reservoirs, moving beyond traditional metrics of energy or power. It opens a new pathway for predictive, mechanism-informed co-optimization of both structure and usage. Whether for fast-charging electric vehicles, ultra-long-life stationary storage, or form-factor-constrained consumer electronics, this reservoir-based framework enables battery design that is not only energy dense — but also engineered for resilience, precision, and real-world endurance across their functional lifetime.

\section*{Governing Equations and Simulation Framework}

\subsection*{Coupled Degradation Model}

This work employs a previously developed physics-based degradation model originally introduced by O'Kane et al.\cite{o2022lithium} and further extended by Ruihe Li et al.\cite{li2023lithium}. The underlying electrochemical foundation is the Doyle--Fuller--Newman (DFN) pseudo-2D model for lithium-ion batteries, which captures charge transport and interfacial reactions across the electrodes and electrolyte. On top of this base model, five key degradation mechanisms are coupled and simulated concurrently within the negative electrode, as implemented in the framework of O'Kane and Ruihe et al.\cite{o2022lithium, li2023lithium}:

\begin{enumerate}
\item SEI Growth: The formation and growth of the solid electrolyte interphase (SEI) layer on the anode irreversibly consumes cyclable lithium. O'Kane et al.\cite{o2022lithium} modeled SEI growth as a diffusion-limited process to capture typical capacity fade behavior. Ruihe et al.\cite{li2023lithium} later implemented a more advanced interstitial-diffusion limited SEI model, which introduces temperature and state-of-charge dependence\cite{li2023lithium}. This work adopts their implementation directly. Additionally, SEI growth occurs not only on the pristine particle surface, but also on newly exposed surfaces due to mechanical cracking---SEI on cracks---which accelerates lithium loss. SEI formation also generates solid by-products that accumulate in the electrode, reducing local porosity. This porosity is treated as a finite design reservoir: its depletion impairs ion transport and accelerates degradation.

\item Lithium Plating: O'Kane et al.'s\cite{o2022lithium} partially reversible plating model is used, where lithium deposits on the anode during charging and is only partially stripped during discharge. Some lithium becomes inactive "dead lithium," leading to permanent loss of lithium inventory. The plating and SEI models are coupled---SEI thickness influences overpotentials and plating reversibility.

\item Particle Cracking: Volume changes in active particles during cycling lead to stress build-up and cracking, modeled via the stress-based mechanics model of O'Kane et al.\cite{o2022lithium} Cracking increases the electrochemically active surface area, thereby enhancing SEI and plating reactions. The SEI on cracks model captures the enhanced reactivity of these fresh surfaces.

\item Stress-Induced Loss of Active Material (LAM): Cracking can also isolate active material, rendering it electrochemically inactive. A critical stress/strain criterion, as proposed by O'Kane et al.\cite{o2022lithium}, determines when active material is lost. Cracking and LAM are coupled through shared mechanical stress evolution.

\item Solvent Dry-Out: To simulate electrolyte consumption, we use the solvent loss model introduced by Ruihe Li et al.\cite{li2023lithium}, which avoids stiff differential equations by implementing solvent depletion in discrete updates. After a pre-defined number of cycles, the model estimates solvent consumption from parasitic reactions and reduces the available electrolyte volume accordingly. This loss of electrolyte acts as an additional degradation pathway---often referred to as electrolyte drying---which increases resistance and limits future SEI/plating activity. Since SEI growth consumes electrolyte solvent, it directly drives this mechanism, linking it to other degradation pathways. In this work, it is assumed there is no external electrolyte reservoir, making solvent loss irreversible.
\end{enumerate}

By incorporating all five degradation mechanisms and their interactions, the model simulates the complex, nonlinear aging behavior observed in lithium-ion cells. For example, the loss of porosity and active surface area amplifies local current density, which in turn accelerates further SEI growth, plating, and cracking. These feedback paths, as demonstrated by O'Kane et al.\cite{o2022lithium}, are essential to reproducing usage-dependent and path-dependent degradation trajectories. All model parameters and couplings used in this work are as reported by O'Kane et al. and Ruihe et al\cite{o2022lithium, li2023lithium}.

\subsection*{Model Implementation and Validation}

The coupled degradation model used in this study is implemented using PyBaMM (Python Battery Mathematical Modeling)\cite{sulzer2021python}, an open-source framework for physics-based lithium-ion battery simulations. We adopt the established DFN-based degradation model introduced by O'Kane et al.\cite{o2022lithium}, with extensions by Li et al.\cite{li2023lithium} to include electrolyte solvent depletion. PyBaMM provides the necessary numerical infrastructure to solve the full set of coupled differential equations governing electrochemical performance, mechanical degradation, and parasitic side reactions, such as SEI growth and lithium plating.

This model has previously been validated by Li et al\cite{li2023lithium}. with experimental data, using degradation mode analysis (DMA), which quantifies individual degradation modes rather than relying solely on aggregate metrics such as capacity fade. In their validation, Li et al.\cite{li2023lithium, li2025importance} evaluated (i) loss of lithium inventory (LLI), primarily due to SEI formation and irreversible lithium plating, and (ii) loss of active material (LAM) in both electrodes, resulting from particle cracking and intra-electrode electrical disconnection. These model predictions were benchmarked against experimentally derived values obtained through open-circuit voltage (OCV) fitting on aged cells.

To contextualize the operating regime explored in this study, we summarize the experimental ageing data referenced during model validation. The data were generated using commercial LG M50T 21700 lithium-ion cells, which feature a high-energy-density design with an NMC811 cathode and a graphite--SiOx composite anode. These 5 Ah cells were tested under controlled thermal conditions using base-cooling at temperatures of $10^\circ$C, $25^\circ$C, and $40^\circ$C. At beginning of life (BoL), the average C/10 discharge capacity was approximately 4.86 Ah.

The cells were cycled within a 70--85\% state-of-charge (SoC) window---a regime intended to isolate SEI growth and electrolyte dry-out as the dominant degradation mechanisms, while avoiding lithium plating and mechanical overstrain. Each ageing cycle involved a 0.3C constant-current--constant-voltage (CC-CV) charge to 4.2 V, followed by a 730 mAh constant-current discharge at 1C (equivalent to 15\% of BoL capacity). Rest steps were included between charge and discharge to allow for thermal and voltage relaxation. Each ageing set consisted of 515 cycles, with cells tested at three temperatures and duplicated for reproducibility.

Reference Performance Tests (RPTs) were conducted at $25^\circ$C after each ageing set to systematically monitor degradation. Two alternating RPT protocols were used: a long protocol, run after even-numbered sets, that included C/10 and C/2 cycling along with galvanostatic intermittent titration technique (GITT) for detailed resistance analysis; and a short protocol, conducted after odd-numbered sets, that combined C/10 cycling with pulse-under-load (PUL) tests to capture dynamic resistance evolution. Alternating these protocols balanced diagnostic resolution with throughput, enabling robust tracking of capacity loss, resistance growth, and mode-specific degradation such as LAM and LLI. Further experimental details and data are available in Kirkaldy et al\cite{kirkaldy2024lithium}.

\subsection*{Modeling the Reservoirs}

The five coupled degradation, experimentally validated model is used to analyze the impact of different reservoirs and determine how they can be tuned to achieve targeted battery performance.

\subsubsection*{Lithium Reservoir}

Cyclable lithium inventory is a critical design reservoir in lithium-ion cells, as its depletion directly leads to capacity loss and performance degradation. In this work, we explore the effect of tuning the lithium reservoir---by adding or removing lithium---on cell performance and degradation outcomes.

Adding excess lithium extends the lithiation capacity of the negative electrode and effectively shifts its open-circuit potential. In this model, excess lithium is incorporated as an additional capacity ($Q_{extra}\ [Ah],\ Eq.\ 1)$, converted to moles of lithium and applied to the negative electrode's initial lithium content. This virtual reservoir is gradually depleted by side reactions such as SEI formation, plating, and stress-induced degradation.

Several experimental strategies support the practical feasibility of lithium reservoir tuning. Pre-lithiation methods, such as using sacrificial lithium-rich additives (e.g., Li$_3$N, Li$_5$FeO$_4$) on the cathode, release lithium during initial charging and compensate for irreversible losses due to SEI growth \cite{su2023review, yang2023roll}. Stabilized lithium metal powder (SLMP) is another widely adopted method, wherein lithium metal powder is applied to the anode surface and activated during assembly, significantly improving initial Coulombic efficiency \cite{qiao2023solid}. Similarly, ex situ pre-lithiated anodes, prepared via electrochemical lithiation or contact with lithium foil, have been shown to preserve lithium inventory and reduce first-cycle losses \cite{he2023regulating}.

More recent innovations include lithium-rich cathodes that inherently contain surplus lithium and in situ lithium replenishment using embedded lithium reservoirs or slow-release additives \cite{liu2024controllable}. These methods have demonstrated improved capacity retention and prolonged cycle life, reinforcing the plausibility of lithium reservoir tuning in practical cells.

In this study, we abstract lithium-consuming degradation processes (such as SEI growth and lithium plating) through the concept of a finite lithium reservoir, and analyze how varying its initial size influences degradation pathways and overall cell longevity. The added lithium is quantified in terms of additional charge capacity ($Q_{extra}\ [Ah])$ and converted to moles using:
\[n_{Li,extra}=\frac{Q_{extra}*3600}{F}\]
where ${F}$ is the Faraday constant. The additional lithium alters the lithium concentration within the negative electrode, calculated as:
\[\Delta c_s=\frac{n_{Li,extra~}}{{\varepsilon }_sL_{NE}A_{NE}}\]
which ultimately shifts the negative electrode's stoichiometry:
\[x'_{NE}=x_{NE}+\frac{\Delta c_s}{c_{s.max}}\]
In our simulations, we use the same cell and negative electrode OCV profile reported by Chen et al.\cite{chen2020development}, who fitted the stoichiometric range as 0.0279 to 0.9014. A 7.5\% increase in cyclable lithium (5\% additional capacity) relative to the baseline corresponds to an increase in the minimum stoichiometry from 0.877 to 0.923, and a decrease in the minimum open-circuit potential from 0.94 V to 0.92 V. The added lithium boosts the initial capacity and also compensates for early SEI-induced lithium loss\cite{attia2022knees}. While this tuning enhances early capacity and stability, it pushes the anode into a deeper lithiation regime with lower equilibrium potential---closer to 0 V vs. Li/Li$^+$, where the risk of lithium plating may increase under fast or cold charging conditions\cite{sieg2019fast}.

\subsubsection*{Porosity Reservoir}

Electrode porosity is a critical design parameter that governs ionic conductivity, electrolyte distribution, and electrochemical reaction kinetics. In silicon--graphite anodes, porosity also plays a key role in buffering silicon expansion during lithiation, mitigating mechanical stress, and delaying electrode degradation\cite{profatilova2020impact}. For this reason, we conceptualize porosity as a finite design reservoir and investigate how tuning its initial value affects battery aging and performance.

Porosity can be tuned in two principal ways: by changing the mass loading of active electrode material or by varying the electrode thickness. While adjusting loading directly affects the areal capacity and N/P (negative/positive) ratio, it also disrupts the initial electrochemical balance of the cell \cite{boyce2022exploring}. Therefore, in this work, we preserve the initial charge balance and isolate porosity effects by changing only the electrode thickness, simulating different calendaring conditions \cite{scheffler2022calendering}. This ensures that the mass of active material per unit area remains constant, preserving gravimetric energy density, while volumetric energy density varies due to structural compaction.

To model this porosity tuning, we introduce a thickness ratio ($R_{thickness}$), defined as the ratio of the modified electrode thickness to the baseline value. This affects the electrode's physical parameters (electrode thickness ($L)$ , the active material fraction (${\varepsilon }_s)$ , and the inactive material fraction (${\varepsilon }_{inactive})$ ) as follows:
\[L'=L*R_{thickness}\]
\[{\varepsilon '}_s=\frac{{\varepsilon }_s}{R_{thickness}}\]
\[{\varepsilon '}_{inactive}=\frac{{\varepsilon }_{inactive}}{R_{thickness}}\]
\[\varepsilon =1-{\varepsilon }'_s-{\varepsilon }'_{inactive}\]
This approach maintains the total active material per unit area, while allowing controlled variation in the void fraction---a key reservoir in our degradation-aware design framework. For example, decreasing the thickness ratio to 0.95 reduces the negative electrode thickness from 85.20 $\mu$m to 80.94 $\mu$m, compressing the structure and reducing porosity from 0.222 to 0.181, while increasing the active material volume fraction from 0.750 to 0.789. Conversely, a thickness ratio of 1.05 leads to less compression, raising porosity to 0.220 and lowering the active material fraction to 0.752. These trends are consistent with experimental observations in silicon--graphite electrodes subjected to varying calendaring forces \cite{profatilova2020impact, yourey2023cell}.

While lower porosity improves volumetric energy density and may stabilize electrode expansion, it can introduce ionic transport limitations and increase the risk of electrolyte dry-out and SEI pore clogging during extended cycling---particularly under high-rate or high-temperature conditions \cite{ma2019towards}. Therefore, calendaring-based porosity tuning, with preserved N/P balance, offers a physically grounded and practically relevant method to optimize the porosity reservoir for targeted performance and lifetime.

\subsubsection*{Electrolyte Reservoir}

In lithium-ion cells, the liquid electrolyte is a finite resource that depletes over time, due to side reactions---most notably the formation and continued growth of the solid electrolyte interphase (SEI). As SEI consumes both lithium and solvent, inadequate electrolyte volume can eventually lead to electrolyte dry-out, where certain regions of the cell lose ionic contact, sharply increasing impedance and triggering a cascade of degradation effects such as active material isolation and lithium plating onset. This phenomenon is well-documented as a leading cause of accelerated capacity fade and sudden end-of-life behavior, often observed as a ``knee'' in the capacity-retention curve \cite{fang2021capacity, li2022modelling}.

To model this effect, we define an electrolyte excess ratio ($R_{excess})$ to explicitly represent the initial overfill beyond what is consumed or trapped. The available electrolyte volume in the cell is described as:
\[{V'}_{EC}=V_{EC}-V_{EC,SEI~}+R_{excess}*V_{Reservoir}\]

where $V_{EC}$ is the nominal electrolyte volume, $V_{EC,SEI}$ is the amount consumed during SEI formation, and the term $R_{excess}*V_{Reservoir}$ quantifies the additional buffer electrolyte. This formulation allows us to evaluate how different levels of initial electrolyte overfill impact degradation and longevity, particularly the delay of dry-out events.

In practice, electrolyte volume is tuned via the electrolyte-to-capacity (E/C) ratio, a design metric describing the amount of liquid per ampere-hour of cell capacity. Commercial lithium-ion cells typically operate with lean electrolyte (E/C $\approx$ 1.3--1.5 g Ah$^{-1}$), enough to just saturate the electrode pores. However, studies have shown that increasing this ratio modestly improves cycle life by ensuring ionic pathways remain functional even after extended SEI growth or plating reactions\cite{wu2021recent,li2021low}.

This becomes even more critical in cells with silicon--graphite anodes, where SEI formation is far more aggressive due to silicon's high surface reactivity and significant volume change during cycling. These cells consume electrolyte not only during initial formation, but also during cycling, as cracking of the SEI on expanding Si particles exposes fresh surfaces. Without sufficient electrolyte to buffer this behavior, such cells may fail prematurely\cite{lee2023sio, an2016state}.

From a modeling standpoint, adding an explicit electrolyte reservoir provides insight into these dynamics. Simulation frameworks have shown that, when electrolyte consumption is included, cells with excess initial volume show a pronounced delay in degradation acceleration compared to lean designs. Conversely, models of dry-out events predict sharp increases in internal resistance and loss of electrochemical activity once the pore space can no longer enable electrolyte ionic transport \cite{wu2021recent,li2021low}.

To sum up, the electrolyte reservoir acts as a buffer against solvent-limited degradation. While excess electrolyte may slightly reduce cell energy density, it plays a critical role in maintaining long-term cell performance---especially in Si-containing systems or high-rate operation scenarios. This modeling approach enables a systematic investigation of electrolyte tuning as a design parameter for longevity optimization.

\newpage

\section*{RESOURCE AVAILABILITY}


\subsection*{Lead contact}

Further information and requests for resources, model implementation details, or simulation data should be directed to and will be fulfilled by the lead contact, Mohammed Asheruddin N (mnazeeru@ic.ac.uk)

\subsection*{Materials availability}

This study did not generate new materials.

\subsection*{Data and code availability}

This study did not generate new experimental datasets. All simulation results were produced using the open-source battery modeling framework PyBaMM (https://pybamm.org). Custom simulation scripts and model modifications developed for this work are available from the lead contact upon reasonable request.

\section*{ACKNOWLEDGMENTS}
This work was supported by the Faraday Institution (grant number FIRG084).

\section*{AUTHOR CONTRIBUTIONS}
Conceptualization, S.E.J.O., M.M., G.J.O.;
Methodology, M.A.N., R.L., S.E.J.O., M.M., G.J.O.;
Software, M.A.N., R.L., S.E.J.O.;
Validation, M.A.N., R.L., S.E.J.O.;
Formal analysis, M.A.N., R.L., S.E.J.O., M.M., G.J.O.;
Investigation, M.A.N., R.L.;
Visualization, M.A.N., M.M., G.J.O.;
Writing – original draft, M.A.N.;
Writing – review \& editing, M.M., G.J.O.;
Supervision, M.M., G.J.O.;
Resources, M.M., G.J.O.;
Funding acquisition, M.M., G.J.O.

\section*{DECLARATION OF INTERESTS}

The authors declare no competing interests.








\newpage


\bibliography{references}

\bigskip


\end{document}